\documentclass[preprint,number,sort&compress,times]{elsarticle}
\journal{arXiv}

\let\todayorig\today
\date{October 27, 2017 by R. Imai, ver. 1.0}
\date{November, 2017 by Y. Yamanaka, ver. 2.0}
\date{December 2, 2017 by R. Imai, ver. 3.0}
\date{December 19, 2017 by Y. Yamanaka, ver. 4.0}
\date{December 23, 2017 by R. Imai, ver. 5.0}
\date{December 25, 2017 by R. Imai, ver. 5.1}
\date{January 10, 2018 by Y. Yamanaka, ver. 6}
\date{January 11, 2018 by R. Imai, ver. 7}
\date{January 16, 2018 by Editage, ver. 8}
\date{January 17, 2018 by R. Imai, ver. 8.1} 
\date{January 17, 2018 by Y. Yamanaka, ver. 8.2}
\date{January 17, 2018 by R. Imai, ver. 8.3}
\date{January 18, 2018 by Y. Yamanaka, ver. 8.4}
\date{January 18, 2018 by R. Imai, ver. 8.5} 
\date{March 1, 2018 by R. Imai, ver. 9}
\date{March 7, 2018 by Y. Yamanaka, ver. 10}
\date{March 7, 2018 by R. Imai, ver. 10.1} 
\date{April 23, 2018 by R. Imai, ver. 12}
\date{April 23, 2018 by R. Imai, ver. 12.1}
\date{May 2, 2018 by Y. Yamanaka, ver. 13}
\date{May 16, 2018 by R. Imai, ver. 14}
\date{May 18, 2018 by Y. Yamanaka, ver. 15}
\date{May 22, 2018 by R. Imai, ver. 15.1}
\date{Aug 5, 2018 by R. Imai, ver. 16}
\date{Aug 8, 2018 by Y. Yamanaka, ver. 17}
\date{Aug 8, 2018 by R. Imai, ver. 17.1} 
\date{}
\let\today\todayorig 
\let\today\relax

\usepackage{bm}
\usepackage{color}
\usepackage{hyperref}
\usepackage{amsmath}
\usepackage{pgf}

\usepackage{color}

\usepackage[italicdiff]{physics}

\newcommand{\Order}[1]{\mathcal{O}\pqty{#1}}
\newcommand{\summintdx}{\sum_{m = -J}^J \int \! dx \,}
\newcommand{\psigs}{\psi^{(0)}}

\newcommand{\oH}{\hat{H}}
\newcommand{\oJ}[1]{\hat{J}^{(#1)}}
\newcommand{\oJp}{\hat{J}^{+}}
\newcommand{\oJm}{\hat{J}^{-}}
\newcommand{\oPi}{\hat{\Pi}}
\newcommand{\osgm}[1]{\hat{\sigma}^{(#1)}}
\newcommand{\osgmp}{\hat{\sigma}^{+}}
\newcommand{\osgmm}{\hat{\sigma}^{-}}
\newcommand{\oa}{\hat{a}}

\newcommand{\ox}{\hat{x}}

\newcommand{\oV}{\hat{V}}

\newcommand{\oR}{\hat{R}}

\newcommand{\lambdath}{\lambda_\text{th}}

\newcommand{\dotot}{\hat{\rho}_{\mathrm{tot}}}
\newcommand{\dos}{\hat{\rho}}
\newcommand{\dor}{\dyad{0}_\mathrm{R}}

\newcommand{\scH}{\bar{H}}
\newcommand{\scV}{\bar{V}}
\newcommand{\scE}{\bar{E}}
\newcommand{\scomega}{\bar{\omega}}

\newif\ifSHOWCHANGES
\usepackage{color}
\newcount\ChangesCount
\ifSHOWCHANGES
\long\def\REMOVEDSTRING#1{{\color{gray}{\it #1}}}

\def\IncChangesCount{\advance\ChangesCount by 1}
\long\def\REMOVED#1{\IncChangesCount\REMOVEDSTRING{#1}\relax}

\def\SHOWCHANGES{{\color{red}\bf Number of changes: \number\ChangesCount}}
\else
\long\def\REMOVED#1{\relax}

\def\SHOWCHANGES{\relax}
\fi

\begin{document}

\begin{frontmatter}

\title{Stability of Symmetry Breaking States\\ in Finite-size Dicke Model with Photon Leakage}

\author[eps]{R.~imai\corref{cor1}}
\ead{rimai@asagi.waseda.jp}

\author[eps]{Y.~Yamanaka}
\ead{yamanaka@waseda.jp}

\cortext[cor1]{Corresponding author}

\address[eps]{Department of Electronic and Physical Systems, Waseda University, Tokyo 169-8555, Japan}

\begin{abstract}
We investigate the finite-size Dicke model with photon leakage.
It is shown that the symmetry breaking states,
which are characterized by non-vanishing $\langle \hat{a} \rangle \neq 0$ and correspond to the ground states in the superradiant phase in the thermodynamic limit, are stable, while the eigenstates of the isolated finite-size Dicke Hamiltonian conserve parity symmetry.
We introduce and analyze an effective master equation that describes the dynamics of a pair of the symmetry breaking states that are the degenerate lowest energy eigenstates in the superradiant region with photon leakage.
It becomes clear that photon leakage is essential to stabilize the symmetry breaking states and to realize the superradiant phase without the thermodynamic limit.
Our theoretical analysis provides an alternative interpretation using the finite-size model to explain results from cold atomic experiments showing superradiance with the symmetry breaking in an optical cavity.
\end{abstract}


\end{frontmatter}


\section{Introduction}
The Dicke model is one of the quantum optical models that has been thoroughly studied  \cite{Dicke1954,Breuer2007openquantum}.
It describes a collection of identical two-level atoms that are coupled with a single electromagnetic mode in a cavity via a dipole interaction.
The significant property of the Dicke model
is that it exhibits a transition from a normal phase to a superradiant phase
when the coupling constant
 takes a critical value in the thermodynamic limit
\cite{Hepp1973-zw,Wang1973,Emary2003-bj,Emary2003-vy}.
Since it is known that this phase transition occurs even at zero temperature,
it is considered to be a quantum phase transition \cite{Emary2003-bj,Emary2003-vy}.
Thanks to recent experimental progress in atomic
 physics, the situations described by the Dicke model have been realized
in cold atomic systems in an optical cavity \cite{Baumann2010-of, Baumann2011-fk},
where the collection of cold atoms plays the same role as a collection of
two-level atoms.
In these experiments the transition to a superradiant
phase is verified by detecting photons leaking from the cavity.
However, this transition cannot be identified to be a quantum transition and/or thermal transition that is defined in equilibrium infinite-size systems, since in the cold atom experiments the number of atoms is finite and the system is open.
Instead, it is suggested that the transition in the experiments can be interpreted as a nonequilibrium phase transition \cite{Dimer2007,Nagy2010,Baumann2010-of}, the photon leaking being taken into account in the thermodynamic limit.
To our best knowledge, this dissipative Dicke model has been investigated in a semi-classical and (plus) stochastic approach based on the thermodynamic limit.
The semi-classical approach \cite{Baumann2010-of,Brennecke2013} implies that quantum operators are replaced with c-numbers and that the superradiant transition is described as a bifurcation of the classical solution. We point out that this treatment ignores quantum fluctuation, {\it i.e.}, the quantum entanglement between the atoms and the cavity mode.
As the stochastic method \cite{Brennecke2013}, a stochastic term that represents a dissipation is added to the Heisenberg equation in such a phenomenological manner that the stochastic operator of the bosonic quasi-particle defined in each of normal and superradiant phases is introduced. Strictly speaking, this quasi-particle picture is exact only in the thermodynamic limit, where the Hamiltonian becomes a corresponding quadratic form. Thus this stochastic method is valid only when the system is close to the thermodynamic limit. When a finite-size system is under consideration instead of the thermodynamic limit, the higher order terms in the Hamiltonian that were neglected in the thermodynamic limit may affect the quasi-particle picture and the gap in theoretical treatment depending on whether the phase is either normal or superradiant is unfavorable. Hence, it is still not clear how the superradiant transition is explained in the finite-size model with the dissipation. Our analysis in this paper focuses on a stability of the superradiant state, without relying on the thermodynamic limit and the quasi-particle picture established then and taking account of quantum fluctuations properly.

For the isolated finite-size Dicke model, there are some previous studies on the singularity in the ground state energy associated with the superradiant phase transition or its finite-size corrections \cite{Vidal2006,Chen2008,Castanos2011}.
There the model is treated as an isolated system without symmetry breaking.
In distinction from the previous research, we introduce the interaction of the finite-size Dicke system with an environment in this paper and focus on the mechanism to realize the symmetry breaking state, which is characterized by $\ev{\oa} \neq 0$, as observed in the experiments \cite{Baumann2010-of,Baumann2011-fk}.

The purpose of this paper is to show that the Dicke model with photon leakage exhibits symmetry breaking, even in finite-size systems.
To achieve this, restricting ourselves to small atomic level spacing,
we first study the ground and first excited states of the isolated Dicke model and estimate an energy gap between them,
because the two states form a pair of degenerate states when the superradiant phase is realized.
Then, we introduce photon leakage out of the cavity to an external vacuum and investigate the temporal evolution of the density matrix in the superradiant region.
It will be shown that the symmetry breaking state becomes stable. Thus, photon leakage is crucial for understanding symmetry breaking in a finite-size Dicke system.

This paper is organized as follows.
In Sec.~\ref{sec:symmetry_sr}, we briefly introduce the Dicke model and
compare the finite-size model with the model in the thermodynamic limit.
The lowest energy eigenstates are constructed under a ``polarization condition'', and we have a pair of the two almost degenerate states, breaking the parity symmetry, in the superradiant region in Sec.~\ref{sec:pair_superradiant}.
Section \ref{sec:sb_state_dynamics} shows that, although the symmetry breaking states are not exact eigenstates, they freeze dynamically.
Considering the leakage photon, we derive and analyze an effective master equation for the two symmetry breaking states in the open Dicke model that interacts with the environment in Sec.~\ref{sec:interaction_with_environment} and discuss the stability of the symmetry breaking states.

\section{Parity symmetry and superradiance}
\label{sec:symmetry_sr}

The Dicke Hamiltonian is given by \cite{Hepp1973-zw,Wang1973,Emary2003-bj,Emary2003-vy}
\begin{align}
\oH_{\rm DH} = \omega_0 \oJ{3} + \omega \oa^\dagger \oa + \lambda \varphi_{\rm c}
\bqty{
    \pqty{\oa + \oa^\dagger} \pqty{\oJp + \oJm}
} \,,
\label{eq:DH0}
\end{align}
where $\oa$ denotes the bosonic annihilation operator for the cavity mode
with frequency $\omega$, and $\hat{J}^{(i)}$ $(i=1,2,3)$ are pseudospin operators describing
a collection of $N$ identical two-level atoms with level spacing $\omega_0$;  these operators obey angular momentum algebra.
The ${\hat J}^\pm$ operators are defined by $\hat{J}^{(1)}\pm i \hat{J}^{(2)}$,
and, we take  $J = N / 2$ for the length of the pseudospin $J$.
The coefficients $\omega_0, \omega$, and $\lambda$ are non-negative.
The symbol $\varphi_{\rm c}$ stands for the normalization factor of wave function for cavity mode \cite{Hepp1973-zw}, namely $\varphi_{\rm c}= 1/\sqrt{2J}$.
Note that we do not neglect the counter-rotating contributions
in the Hamiltonian (\ref{eq:DH0}).
As in Ref.~\cite{Chen2008}, we employ the following Hamiltonian,
transformed by the unitary operator ${\hat U}
=\exp [i (\pi / 2)\oJ{2}]$,
\begin{align}
    \oH &={\hat U}\oH_{\rm DH}  {\hat U}^\dagger
    =- \omega_0 \oJ{1} + \omega  \oa^\dagger \oa
    + 2 \lambda \varphi_\mathrm{c} \pqty{\oa + \oa^\dagger} \oJ{3} \,,
\end{align}
because the diagonalized form of the interaction term is convenient for our arguments.
The parity transformation in this representation is executed by the unitary operator
\begin{align}
    \oPi = \exp \bqty{i \pqty{\oa^\dagger \oa - \oJ{1}} \pi}\,,
\end{align}
and $\oH$ is invariant under the parity transformation $\oPi$, namely $[\oH, \oPi] = 0$.

In the thermodynamic limit, where $N \to \infty$,
the system shows two phases that separated by the critical coupling constant
$\lambda_c = \sqrt{\omega_0 \omega}/2$ \cite{Emary2003-bj,Emary2003-vy}.
For $\lambda < \lambda_c$, the system is in the normal phase, where the eigenstates are symmetric, that is, they each have a definite parity.
For $\lambda > \lambda_c$, the system is in the superradiant phase where the atoms are collectively excited and
the light field obtains a coherent amplitude.
The ground state in the superradiant phase breaks the parity symmetry \cite{Emary2003-vy}, which means that the generation of the superradiant phase is interpreted as a spontaneous symmetry breaking with the nonvanishing order parameter $\ev{\oa}\neq 0$.

In the finite-size model that we will focus on, the ground and first excited state form a degenerate pair in the superradiant region \cite{Castanos2011,Larson2017}, which is characterized by closing the energy gap between them.
It is also reported that the similar pair formation presents in the higher excited states \cite{Larson2017,Ricardo2013}.
These degeneracies occur asymptotically as $\lambda$ increases.
It is, as will be shown, essential to form symmetry breaking states in the superradiant region.

\section{Formation of a degenerate pair in superradiant region}
\label{sec:pair_superradiant}

To investigate a degenerate pair with the two lowest states analytically, we will construct the ground and first excited states in a perturbative manner.
For convenience, we introduce the scaled Hamiltonian $\scH$,
\begin{align}
    \scH = - \scomega_0 \oJ{1} + \scomega \oa^\dagger \oa + \pqty{\oa + \oa^\dagger} \oJ{3}\,,
\end{align}
where
\begin{align}
    \scomega_0 = \frac{\omega_0}{2 \lambda \varphi_\mathrm{c}}\,,
    \quad
    \scomega = \frac{\omega}{2 \lambda \varphi_\mathrm{c}}\,.
\end{align}
First, we consider the limiting case of $\scomega_0 = 0$, keeping $\scomega$ finite.
In this limit, we can construct all the eigenstates in the following way.
Since $\comm*{\scH}{\oJ{3}} = 0$, we can represent $\scH$ in each subspace,
labeled by the eigenvalue $m$ of $\oJ{3}$, as
\begin{align}
    \scH_m =
    \scomega \oa_m^\dagger \oa_m
    - \frac{m^2}{\scomega}\, ,
\end{align}
where
$
    \oa_m = \oa + d_m,~ d_m = m / \scomega
$\,.
Then, the eigenstates of $\scH_m$ are exhausted by
$\ket{m} \otimes \pqty{\oa_m^\dagger}^n \ket{0_m}/\sqrt{n!}$,
where the coherent state $\ket{0_m}$ is defined by
\[
    \oa \ket{0_m} = - d_m \ket{0_m}\,.
\]
Obviously, there are two orthogonal ground states, $\ket{\Psi_0^{(\pm J)}}$,
\begin{align}
    \ket{\Psi_0^{(\pm J)}} &= \ket{\pm J} \otimes \ket{0_{\pm J}}.
\end{align}
This ensures ground state degeneracy in our limiting case.

Next, we revive the $\scomega_0$ term while restricting ourselves to the assumption $0 < \scomega_0 \ll 1$.
Then, the $\scomega_0$ term can be regarded as a perturbation, which lifts the degeneracy of the two ground states at $\scomega_0 = 0$ at the $2J$ th-order.
Since the non-degenerate eigenstate has well-defined parity, we have the ground state $+1$ (even) parity, $\ket{\Psi_0}$ and the first excited state $-1$ (odd) parity, $\ket{\Psi_1}$.
Hence, they are given by
\begin{align}
    &\ket{\Psi_0} = \frac{1}{\sqrt{2}} \Big[ \ket{\Psi_0^{(-J)}} + \ket{\Psi_0^{(+J)}} \Big]
        + \Order{\delta}\,,
        \label{eq:gs}\\
    &\ket{\Psi_1} = \frac{1}{\sqrt{2}} \Big[ \ket{\Psi_0^{(-J)}} - \ket{\Psi_0^{(+J)}} \Big]
        + \Order{\delta}\,,
        \label{eq:1st}
\end{align}
since $\oPi \ket{\Psi_0^{(\pm J)}}=\ket{\Psi_0^{(\mp J)}}$.
Here, $\delta$ is estimated as $\delta = \scomega_0 \scomega / \sqrt{N} = \sqrt{N} \lambda_c^2 / \lambda^2$ in the leading order.
For $\delta \ll 1$, or $\lambda \gg \lambda_c N^{1/4}$, which we will call polarization condition, the terms $\Order{\delta}$ in the expressions \eqref{eq:gs} and \eqref{eq:1st} are negligible.

As reported in the previous researches \cite{Castanos2011,Larson2017}, the energy gap between the ground and first
excited states $\Delta E = \mel{\Psi_1}{\oH}{\Psi_1} - \mel{\Psi_0}{\oH}{\Psi_0}$ closes to zero even in the superradiant region of the finite-size model.
Here, we shall clarify this asymptotic degeneracy through perturbative evaluation.
Let us define the position operator for $\oa$ by $\ox = (\oa + \oa^\dagger)/\sqrt{2\omega}$.
Then, a state of the Dicke model is generally represented as
\begin{align}
    \ket{\Psi} = \summintdx \psi_m(x) \ket{m,x} \,,
\end{align}
where a basis $\ket{m,x}$ is a common eigenvector of both $\oJ{3}$ and $\ox$ with eigenvalue $m$ and $x$, respectively.
Note that the transformation of the basis by parity operator $\oPi$ is represented by $\Pi \ket{m,x} = \ket{-m,-x}$.
Since the parity of the ground state is even, namely $\oPi \ket{\Psi_0} = \ket{\Psi_0}$.
The ground state can be represented in the position basis as
\begin{align}
    \ket{\Psi_0} = \summintdx \psigs_m (x) \ket{m,x} \,,
\end{align}
with non-negative coefficients $\psigs_m(x)$.
Next, using the ground state coefficients, we define a twisted trial state for the first excited state,
\begin{align}
    \ket{\Psi'} = \sum_{m > 0} \int dx \, \psigs_m (x) \ket{m, x} - \sum_{m < 0} \int dx \, \psigs_m (x) \ket{m, x}.
\end{align}
Here we assume $J$ to be half-integer for simplicity. However, the result below \eqref{eq:upper_bound_E} holds even for integer $J$.
The energy difference $\Delta E' = \mel{\Psi'}{\oH}{\Psi'} - \mel{\Psi_0}{\oH}{\Psi_0}$ is evaluated as
\begin{align}
    \Delta E' = \omega_0 (2J + 1) \int dx \, \psigs_{\frac{1}{2}} (x) \psigs_{-\frac{1}{2}} (x)\,.
\end{align}
The true energy gap $\Delta E$ is smaller than $\Delta E'$ since the trial state is an odd-parity state and orthogonal to the ground state.
Form the perturbative calculation presented in
\ref{appendix:calc_upper_bound}, we have upper bounds
on $\psigs_{\pm \frac{1}{2}} (x)$, \eqref{eq:psi_upper_bound} and
\eqref{eq:psi_upper_bound2}.
In conclusion, we find that $\Delta E$ is bounded as
\begin{align}
    \Delta E < \Delta E' < \omega_0 (N + 1) \pqty{\frac{\lambdath}{\lambda}}^{N - 1}
    \label{eq:upper_bound_E}
\end{align}
with a threshold $\lambdath = {\sqrt{e \omega_0 \omega / 2}} = \sqrt{2e} \, \lambda_c$,
where $e$ is Napier's constant.
When $\lambda > \lambdath$, which we will call superradiant region, this bound shows that
the energy gap approaches zero exponentially as $N$ increases, and that the ground and first excited state form a degenerate pair.
On the other hand, when $\lambda < \lambdath$, the right-hand side of \eqref{eq:upper_bound_E} grows as $N$ increases, and the inequality has no
critical bound, which means the energy gap may remain finite even in the thermodynamic limit.

\section{Realization of a state with broken parity--symmetry and frozen dynamics}
\label{sec:sb_state_dynamics}

Using the ground and first excited states, $\ket{\Psi_0}$ and
$\ket{\Psi_1}$ in the previous section, coherent superposition states can be defined by
\begin{align}
    &\ket{+} = \frac{1}{\sqrt{2}} \bqty{ \ket{\Psi_0} + \ket{\Psi_1} }
    = \ket{\Psi_0^{(-J)}} + \Order{\delta} \,, \\
    &\ket{-} = \frac{1}{\sqrt{2}} \bqty{ \ket{\Psi_0} - \ket{\Psi_1} }
    = \ket{\Psi_0^{(+J)}} + \Order{\delta}\,,
\end{align}
which exhibit cavity mode coherence,
\begin{align}
    \oa \ket{\pm} = \pm \alpha \ket{\pm} + \Order{\delta}\,,
    \label{eq:a_coherent}
\end{align}
where $\alpha = J/\scomega$.
In the experiment \cite{Baumann2011-fk},
the coherent cavity mode in the superradiant
phase is observed either in $\phi=0$ or in $\phi=\pi$,
where $\phi$ is the relative time phase with the coupling laser, and we may
interpret these two observed states as $\ket{\pm}$.
However, since they are not eigenstates of the Hamiltonian, they cannot be stationary.
Suppose that the system is initially (say, at $t=0$) prepared in $\ket{+}$. The
temporal evolution of the system is analytically expressed as
\begin{align*}
\abs{\braket{+}{\Psi(t)}}^2 = \cos^2 \frac{\Delta E t}{2}\,
,\quad
\abs{\braket{-}{\Psi(t)}}^2 = \sin^2 \frac{\Delta E t}{2}\,,
\end{align*}
where we have ignored contributions of the $\delta$ order, assuming that the polarization condition is satisfied.
These solutions show that the system oscillates
 between $\ket{+}$ and $\ket{-}$ states with period $T = 2 \pi/\Delta E$.
As shown in Sec.~\ref{sec:pair_superradiant}, in the superradiant region,
as the number of atoms increases, the gap $\Delta E$ approaches zero. Consequently,
the oscillation freezes over experimental time scales, which is an essential prerequisite for observing symmetry breaking, as will be seen below.

\section{Interaction with environment}
\label{sec:interaction_with_environment}

In the previous section, we have examined the possibility that a state
 represented by an arbitrary linear combination of
$\ket{\Psi_0}$ and $\ket{\Psi_1}$ freezes.
This does not explain the experimental results \cite{Baumann2011-fk} that either of the symmetry breaking state $\ket{+}$ or $\ket{-}$ is observed.
Here, we take account of the experimental situation where the system interacts with the environment.
For simplicity, we will discuss the stability of the low-energy states under the interaction with the environment.
To achieve this, we suppose that the Dicke system is initially in a state represented in any linear combination of the ground and first excited states.
As is seen below, once the Dicke system is in the subspace of the two-lowest energy eigenstates, the dissipative temporal evolution with the environment is confined in the subspace effectively.
We will show that a dissipation associated with the open system is a crucial element for realizing the symmetry breaking state in the superradiant region.

We define the Hamiltonian for the total system by adding the free photon reservoir to the Dicke Hamiltonian
with the coupling between the cavity mode and the environment (reservoir)
modes:
\begin{align}
    \oH_\mathrm{tot} = \oH + \oH_\mathrm{R} + \oV \,,
\end{align}
where $\oH_\mathrm{R}$ and $\oV$ denote the free part of the reservoir and the coupling term, respectively.
They are given by
\begin{align}
    \oH_\mathrm{R} &= \int_{-\infty}^\infty dk \, \Omega_k \oR^\dagger_k \oR_k\,,
    \\
    \oV &= \int_{-\infty}^\infty dk \, g_k \pqty{ \oa + \oa^\dagger } \pqty{ \oR_k + \oR_k^\dagger }\,,
\end{align}
where $\Omega_k = c_\mathrm{L} \abs{k}$ with speed of the light $c_\mathrm{L}$.
The operator $\oR_k$ is a bosonic operator for wave number $k$ mode in the reservoir.
We consider the situation in which the reservoir is initially in the vacuum state,
namely $\oR_k \ket{0}_R = 0$, and the coupling strength $g_k$ is infinitesimal.
To derive a master equation for the system, we employ the Born approximation
\cite{Breuer2007openquantum,QuamtumMeasurementAndControl2009},
where correlations between the system and the reservoir, which appears in the temporal evolution, is assumed to be sufficiently small and that the state of the reservoir remains in the vacuum.
In this approximation, the density operator of the total system can be represented by a product state of the partial density operators for the system and the reservoir, namely
\begin{align}
    \dotot(t) = \dos(t) \otimes \dor\,.
\end{align}
In the interaction picture, in which the temporal evolution for a operator $O$ is defined by $O (t) = e^{i \oH_0 t} O e^{-i \oH_0 t}$ where $\oH_0 = \oH + \oH_\mathrm{R}$, the temporal evolution of the reduced density operator for the Dicke system can be described by the quantum master equation \cite{Breuer2007openquantum,QuamtumMeasurementAndControl2009}
\begin{align}
    \dv{t} \dos(t) = - \int_0^t dt' \, \tr_\mathrm{R} \comm{\oV(t)}{\comm{\oV(t')}{\dos(t') \otimes \dor}}\,,
    \label{eq:qme_born_approx}
\end{align}
where $\tr_\mathrm{R}$ denotes a partial trace over the Hilbert space
of the reservoir.
When the polarization condition is satisfied, or $\delta \ll 1$, as discussed in the Sec.~\ref{sec:pair_superradiant}, the symmetric eigenstates are represented by
Eqs.~\eqref{eq:gs} and \eqref{eq:1st}.
Noting relations
\begin{align*}
    \oa \ket{\Psi_0} &= \alpha \ket{\Psi_1} + \Order{\delta} \,,\\
    \oa \ket{\Psi_1} &= \alpha \ket{\Psi_0} + \Order{\delta} \,,
\end{align*}
we find that the cavity mode $\oa$ in the interaction picture is represented in the basis of $\boldsymbol{\Psi} = (\ket{\Psi_0}~ \ket{\Psi_1}~ \ket{\Psi_2}~ \dots)$ as
\begin{align}
    \oa(t) = \boldsymbol{\Psi}
    \left(
    \begin{array}{c|c}
        \begin{matrix}
            0 & \alpha e^{i \Delta E t} \\ \alpha e^{-i \Delta E t} & 0
        \end{matrix} & \Order{\delta} \\
        \hline
        \Order{\delta} & \ddots
    \end{array}
    \right)
    \boldsymbol{\Psi}^{\dagger}
    \label{eq:a_matrix}
\end{align}
where $\Delta E$ denotes the energy gap between the ground and the first excited states as defined in Sec.~\ref{sec:symmetry_sr}.
Thus the transition matrix elements of $\oa$ between the subspace spanned by
$\ket{\Psi_0}$ and $\ket{\Psi_1}$ and the rest of the Hilbert space is suppressed on the order of $\delta$.
We ignore $\Order{\delta}$ in Eq.~\eqref{eq:a_matrix}, and may restrict ourselves to the subspace of $\ket{\Psi_0}$ and $\ket{\Psi_1}$ that we are interested in if we assume that
state of the Dicke system initially belongs to the subspace.
Namely, the complementary subspace of states with higher energies can be eliminated even if the interaction with the environment exists.
Then, we can represent $\oa$ in the subspace with a two-state operator $\osgmp = \dyad{\Psi_1}{\Psi_0} = (\osgmm)^\dagger$ as
\begin{align}
    \oa(t) = \alpha e^{i \Delta E t} \osgmm + \alpha  e^{-i \Delta E t} \osgmp \,.
    \label{eq:a_in_the_subspace}
\end{align}
From Eqs. \eqref{eq:derv_qme1} and \eqref{eq:a_in_the_subspace}, the temporal behavior of the state in the two lowest energy eigenspace $\dos_2$ is described by
\begin{multline}
    \dv{t} \dos_2 (t) = - 4 \alpha^2 \int_0^t dt' F(t-t') \\
    \times
    \Big\{ e^{i \Delta E (t - t')} \bqty{\osgmm \osgmp \dos_2 (t') - \osgmm \dos_2 (t') \osgmp}
    \\
    + e^{-i \Delta E (t - t')} \bqty{\osgmp \osgmm \dos_2 (t') - \osgmp \dos_2 (t') \osgmm}
    \\
    - e^{i \Delta E (t + t')} \osgmm \dos_2 (t') \osgmm - e^{-i \Delta E (t + t')} \osgmp \dos_2 (t') \osgmp
    \Big\} + (\textrm{h.c.})\,,
    \label{eq:derv_qme1}
\end{multline}
where
\begin{align}
    F(t) = \int_{-\infty}^{\infty} dk \, g_k^2 e^{-i \Omega_k t}\,.
    \label{eq:F}
\end{align}
As seen in Sec.~\ref{sec:pair_superradiant}, the energy gap $\Delta E$ closes to zero as the number of atoms increases in the superradiant region.
Therefore, here we substitute 0 for $\Delta E$ in Eq.~\eqref{eq:derv_qme1} to obtain
\begin{align}
    \dv{t} \dos_2(t)  = 8 \alpha^2 \int_0^t dt' \, \Re F(t-t') \Big\{ \osgm{1} \dos_2(t') \osgm{1} - \dos_2(t') \Big\}\,,
\end{align}
where we defined $\osgm{1} = \osgmp + \osgmm = \dyad{+} - \dyad{-}$.
Using Cauchy principal value, we can evaluate the integral in Eq.~\eqref{eq:F} as \cite{Moy1999}
\begin{align}
    F(t)
    = -2ig^2 \mathcal{P} \frac{1}{t} + 2\pi g^2 \delta(t)\,,
    \label{eq:F_PV}
\end{align}
where the coupling strength between the cavity mode and the reservoir is assumed to be independent of $k$, and is denoted by $g$.
Since the real part of $F(t)$ is given as the delta function, the quantum master equation becomes a Markovian form
\begin{align}
    \dv{t} \dos_2(t)  = \frac{\gamma}{2} \Big\{ \osgm{1} \dos_2(t) \osgm{1} - \dos_2(t) \Big\}\,,
\end{align}
where $\gamma = 32 \pi \alpha^2 g^2$ is used.
This is an effective master equation for the degenerate pair in the superradiant region.
It is worth noting that the closing of the energy gap yields the Markovian equation without making use of the Markov approximation \cite{Breuer2007openquantum,QuamtumMeasurementAndControl2009}.
In a matrix representation with the notation $\rho_{\alpha \beta} = \bra{\alpha} \dos_2 \ket{\beta}$,
it is given as
\begin{align}
    \dv{t}
    \begin{pmatrix}
    \rho_{++}(t) & \rho_{+-}(t) \\
    \rho_{-+}(t) & \rho_{--}(t)
    \end{pmatrix}
    = - \gamma
    \begin{pmatrix}
    0 & \rho_{+-} \\
    \rho_{-+} & 0
    \end{pmatrix}\,.
    \label{eq:qme_mel}
\end{align}

For example, if the system is initially in the symmetric eigenstates of $\oH$ given in Eqs.~\eqref{eq:gs} and \eqref{eq:1st}, the density operators $\dyad{\Psi_0}$ and $\dyad{\Psi_1}$ are represented in the matrix form as
\begin{align}
    \begin{pmatrix}
        1/2 & 1/2 \\
        1/2 & 1/2
    \end{pmatrix}
    \quad &\text{(ground state)}\,, \label{eq:rhoinig}\\
    \begin{pmatrix}
    1/2 & -1/2 \\
    -1/2 & 1/2
    \end{pmatrix}
    \quad &\text{(first excited state)}\label{eq:rhoinie}\,,
\end{align}
then the dephasing occurs in the temporal evolution, which means that the off-diagonal elements exponentially decay to zero, and that diagonal ones only survive without change.
Without the photon leakage, only the frozen dynamics in Sec.~\ref{sec:sb_state_dynamics} is seen.
The photon leakage gives rise to the dephasing process.
Finally, the density operator with an initial condition of
any linear combination of Eqs.~\eqref{eq:rhoinig} and \eqref{eq:rhoinie} reaches
 the maximally mixed state
\begin{align}
    \begin{pmatrix}
    1/2 & 0 \\
    0 & 1/2
    \end{pmatrix}\,.
\end{align}
This result shows that the eigenstates of $\oH$  are fragile and collapse into $\ket{\pm}$.
In other words, $\ket{\pm}$ are the only stable pure states.
It is notable that the photon leakage changes the stable states of the system from the symmetric eigenstates of $\oH$ to the symmetry breaking states $\ket{\pm}$ with changing the expectation of $\oa$ form $\ev{\oa} = 0$ to $\ev{\oa} = \pm \alpha$.
As a result of the appearance of $\ket{\pm}$, coherent states are realized in experiments.
It must be noted that the two-state approximation employed in this section is available when the polarization condition $\delta \ll 1$ is satisfied and that the exact degeneracy on the lowest eigenstate pair, namely $\Delta E = 0$, is assumed.
Since the polarization condition demands far stronger coupling than the critical value, it is not acceptable to extend the above results immediately to the dynamics near the phase boundary.

\section{Summary}
In this study, we have investigated the finite-size Dicke model with photon leakage in the superradiant region.
It has been shown that the symmetry breaking states are stable.
We formulated the effective master equation for the degenerate pair whose solutions show dephasing into the symmetry breaking states even in the finite-size model.
The characteristic property of the stable eigenstates of the isolated finite-size Dicke Hamiltonian, i.e., without photon leakage or dissipation, is that they preserve the parity symmetry exactly.
Then, once photon leakage begins, the symmetry breaking states are instead stabilized.
This suggests that, if there is the photon leakage, the symmetry breaking superradiant state can be realized without the thermodynamic limit.
Although our analysis is on the premise that the polarization condition holds, it gives a full quantum description of the stability of the superradiant state in the dissipative Dicke model without taking the thermodynamic limit explicitly.

Our conclusion is drawn from the analysis under the polarization condition in the superradiant region.
A more general study of the model beyond these bounds is desirable.

\section*{Acknowledgments}
We are grateful to K.~Yuasa and Y.~Nakamura for valuable discussions.
We also thank YITP at Kyoto University and RIKEN iTHES/iTHEMS for offering us the opportunity to discuss this work during the workshops on ``Thermal Quantum Field Theories and Their Applications" (2017) (YITP-W-17-11) and "Frontiers of nonequilibrium physics -- Particle physics, cosmology, and condensed matter --" (2017), respectively.
This work is supported in part by JSPS KAKENHI Grant No.~16K05488.

\appendix
\section{Derivation of an upper bound}
\label{appendix:calc_upper_bound}

The unperturbed part of the Hamiltonian $\scH_0 = \scomega \oa^\dagger \oa + \pqty{\oa + \oa^\dagger} \oJ{3}$ has two degenerate ground states $\ket*{\Psi_0^{(+J)}}$ and $\ket*{\Psi_0^{(-J)}}$. We begin with the projection operators
\[
    P = \dyad*{\Psi_0^{(+J)}} + \dyad*{\Psi_0^{(-J)}}, ~ P + Q = 1.
\]
Let $\scE_0^{(0)}$ and $\scE_0$ be the ground energy of $\scH_0$ and $\scH = \scH_0 + \epsilon \scV$, respectively, with the perturbation defined as $\scV = - \scomega_0 \oJ{1}$.
Inserting the projection operators into the eigenstate equation
$[\scH_0 + \epsilon \scV ] \ket*{\Psi_0} = \scE_0 \ket*{\Psi_0}$, we obtain
\begin{align}
    \begin{split}
        \epsilon P \scV \ket{\chi} &= ( \scE_0 - \scE_0^{(0)} ) \ket{\varphi}, \\
        \epsilon Q \scV \ket{\varphi} &= ( \scE_0 - \scH_0  - \epsilon Q \scV Q ) \ket{\chi}.
    \end{split}
\end{align}
where $\ket{\varphi} = P \ket{\Psi_0}, \ket{\chi} = Q \ket{\Psi_0}$, and we used the relation $P \scV P = 0$. From the definition,
\begin{multline*}
    \psi_{-\frac{1}{2}} (x)
    =
    \bra{- \frac{1}{2}, x}
    \epsilon (\scE_0 - \scH_0 - \epsilon Q \scV Q)^{-1} Q \scV \ket{\varphi}
    \\=
    \bra{- \frac{1}{2}, x}
    \bigg[
    \frac{\epsilon}{\scE_0 - \scH_0} Q \scV
    \hspace{6ex}
    \\+
    \frac{\epsilon}{\scE_0 - \scH_0} Q \scV Q \frac{\epsilon}{\scE_0 - \scH_0} Q \scV + \dots
    \bigg]
    \ket{\varphi}\,.
\end{multline*}
We see that the $\epsilon^{\zeta}$ terms with $\zeta < J-\frac{1}{2}$ disappear.
It is found that the lowest-order contribution is given by
\begin{multline*}
    \psi_{-\frac{1}{2}} (x) \simeq \pqty{\frac{\scomega_0 \epsilon}{2}}^{J - \frac{1}{2}}
    \bra{- \frac{1}{2}, x}
    \frac{1}{\scH_0 - \scE_0} Q \oJp Q
    \\ \cdots \frac{1}{\scH_0 - \scE_0} Q \oJp Q \cdots
    \frac{1}{\scH_0 - \scE_0} Q \oJp
    \ket{\Psi^{(-J)}_0}.
\end{multline*}
This contains a sequence of $Q$, which represents repeated projections onto intermediate states.
Considering low-energy processes, we see the dominant contribution comes from the following intermediate states:
\[
    \ket{\Psi^{(-J + 1)}_0}, \ket{\Psi^{(-J + 2)}_0}, \cdots , \ket{\Psi^{(-\frac{3}{2})}_0}.
\]
Then $\psi_{-\frac{1}{2}} (x)$ can be evaluated as products of matrix elements
\begin{multline}
    \psi_{-\frac{1}{2}} (x) \simeq \pqty{\frac{\scomega_0 \epsilon}{2}}^{J - \frac{1}{2}}
    \bra{- \frac{1}{2}, x}
    \frac{1}{\scH_0 - \scE_0} \ket{\Psi_0^{(-\frac{1}{2})}}
    \bra{\Psi_0^{(-\frac{1}{2})}} \oJp
    \ket{\Psi_0^{(-\frac{3}{2})}}
    \\ \times
    \bra{\Psi_0^{(-\frac{3}{2})}}
    \frac{1}{\scH_0 - \scE_0} \ket{\Psi_0^{(-\frac{3}{2})}}
    \bra{\Psi_0^{(-\frac{3}{2})}} \oJp \ket{\Psi_0^{(-\frac{5}{2})}}
    \\
    \cdots
    \\ \times
    \bra{\Psi_0^{(-J+1)}}
    \frac{1}{\scH_0 - \scE_0} \ket{\Psi_0^{(-J+1)}}
    \bra{\Psi_0^{(-J+1)}} \oJp \ket{\Psi_0^{(-J)}}.
    \label{eq:ap2}
\end{multline}
To evaluate these matrix elements, we employ the following inequalities:
\begin{align*}
    \bra{\Psi_0^{(-m+1)}} \oJp \ket{\Psi_0^{(-m)}} &< J + \frac{1}{2},
    \\
    \prod_{m = -J+1}^{-1/2} \bra{\Psi_0^{(m)}} \frac{1}{\scH_0 - \scE_0} \ket{\Psi_0^{(m)}}
    &= \prod_{m = -J+1}^{-1/2} \frac{\scomega}{(J + m)(J - m)} \\
    &< \bqty{\frac{e \scomega}{(J + \frac{1}{2})(J - \frac{1}{2})}}^{J - \frac{1}{2}}.
\end{align*}
Since $N \ge 2$, we
found that \eqref{eq:ap2} is bounded as
\begin{align}
    \psi_{-\frac{1}{2}} (x) < \pqty{\frac{e \omega_0 \omega}{2 \lambda^2}}^{J - \frac{1}{2}}.
    \label{eq:psi_upper_bound}
\end{align}
Following the same procedure for $\psi_{\frac{1}{2}} (x)$, we obtain
\begin{align}
    \psi_{\frac{1}{2}} (x) < \pqty{\frac{e \omega_0 \omega}{2 \lambda^2}}^{J - \frac{1}{2}}.
    \label{eq:psi_upper_bound2}
\end{align}

\SHOWCHANGES

\section*{References}

\end{document}